# Weak+Vacuum and One Decoy States with Two Way Quantum Key Distribution Protocol

M. F. Abdul Khir[1,2], Iskandar Bahari[3], S. Ali[4], S. Shaari[1]

[1]Photonics Lab, Institute of Micro Engineering and Nanoelectronics (IMEN),
Universiti Kebangsaan Malaysia, 43400 UKM Bangi, Malaysia.
[2]Photonics Lab, Photonic Technology & Product Development,
[3]Advance Analysis and Modelling,
Mimos Berhad, Technology Park Malaysia, 57000 Kuala Lumpur, Malaysia
[4]Department of Science in Engineering, Faculty of Engineering,
International Islamic University of Malaysia (IIUM), Jalan Gombak, 53100 Kuala Lumpur, Malaysia

**Abstract**

We present relevant bounds for the case of weak+vacuum decoy state and one decoy state for a two way four states Quantum Key Distribution (QKD) protocol. The numerical simulation result was significant given that an improvement in maximum secure distance of nearly double is achieved.

## 1. INTRODUCTION

Since the introduction of the first Quantum Key Distribution (QKD) protocol, namely BB84 by Bennet and Brassard in 1984 [1], rapid progress in the field has been recorded, from theoretical down to experimental aspects [2]. In recent progresses, attempts of applying QKD into classical network infrastructure were also reported [3-6]. With its security guaranteed through law of physics, QKD has managed to provide the unconditionally secure method for secret key sharing between two distant parties. However, with the absent of ideal equipments, realization of an information theoretically secure QKD appeared less practical. The unavailability of true single photon source has made most QKD systems rely on weak laser pulses which inevitably emit multi photon pulses, a situation that invites powerful attacks such as Photon Number Splitting (PNS) attack. While the security of the shared key remains unconditionally secure, this limitation severely affects the secure key generation rate and the maximum secure distance.

Fortunately, with recent introduction of decoy state method by Hwang [7] and further development by such as in [8~14], the weak laser pulses based QKD system is made practical again. Since then, applications of decoy state have been demonstrated successfully in several experimental works [15~18]. Recently, progresses in decoy states with regards to QKD protocols other than BB84 have also been reported such as found in the work by [19,20] for SARG04. For two way QKD protocols [21~29], practical decoy state extension can be seen in the work of Shaari et al in [30]. Their result with two decoy states for the LM05 protocol was encouraging given that the secure distance has now been extended by almost double.

In our previous work in [32], we have considered the first approach suggested in [30] where the single and double photon contributions were separately calculated, to accommodate the case of weak+vacuum decoy state. In this work, we simplify the bound in the second approach in [30] where the single and double photon contributions were lumped and also the case for one decoy state with both approaches. We then conduct numerical analysis of the proposed bounds. In order to find out how well the proposed scheme would perform, we also compare against the case of without decoy state and the case of infinite decoy state. As such, this letter is organized as follows. Section two describes the proposed scheme. Section three discusses the results and section four conclude and suggest future works.

## 2. DECOY STATE ESTIMATIONS

### 2.1 Infinite Decoy State

Before we proceed with the two cases of weak+vacuum and one decoy states, let us first review the case of theoretical limit in which we assume that we have an infinite number of decoy states. We assume a two way QKD system with channel transmission $t_C = 10^{-\left(\frac{2\alpha l_C}{10}\right)}$ where $\alpha$ is the optical fiber loss coefficient and $l_c$ is the transmission distance. Notice the factor of two which represents the two way quantum channel of this two way protocol in which we assume the same loss at both forward and backward channel. The overall transmission and detection efficiency is then $\eta = t_C \eta_{Bob}$ and the transmittance of i-th photon state $\eta_i = 1 - (1 - \eta)^i$. The yield $Y_i$, gain $Q_i$ and error rate $E_i$ for i-th photon states is given in [9] as :

$$Y_i = Y_0 + \eta_i - \eta Y_0 \cong Y_0 + \eta_i \tag{1}$$

$$Q_i = Y_i \frac{\mu^i}{i!} e^{-\mu} \tag{2}$$

$$E_i = \frac{e_0 Y_0 + e_{detector} \eta_i}{Q_i} \tag{3}$$

where $Y_0$ is background detection events, $e_0$ is noise error rate, assumed as 1/2 due to randomness and $e_{detector}$ is the erroneous signal detection probability.

The overall Signal Gain ($Q_\mu$) and Quantum Bit Error Rate QBER ($E_\mu$) as given in [9] is then

$$Q_\mu = \sum_{i=0}^{\infty} Y_i \frac{\mu^i}{i!} e^{-\mu} = Y_0 + 1 - e^{-\eta \mu} \tag{4}$$

$$E_\mu = \frac{e_0 Y_0 + e_{detector}(1 - e^{-\eta\mu})}{Q_\mu} \quad (5)$$

## 2.2 Weak+Vacuum Decoy State

The weak+vacuum decoy state was first proposed by Lo et al as the optimal case for a practical two decoy states with one as a vacuum state and the other as a weak state [9]. The key advantage in the weak+vacuum decoy state is that Bob and Alice can estimate the background rate correctly through the vacuum decoy state [9]. This then lead to a better bound and was in fact shown to be optimal for BB84 in [9].

In the course of extending decoy state into a two way protocol, one should also consider the gain from double photon pulses besides the single photon pulses. This was the major consideration in the work of [30] which results into two key rate formulas. While the first, represented in Eq 25 of [30] calculates single and double photon contributions separately, the second, represented in Eq 26 lumped both contributions. We describe the case for the former which is adapted from our previous work in [32] in sub-section A and the latter in sub-section B.

### A. The case for $R_{1+2}$

As previously mentioned, in this case, contribution from single and double photon pulses are calculated separately. For single photon contribution, we can directly use the one suggested by Ma et al in [9] while for double photon contribution we need to derive from the one suggested for the case of two decoy state by Shaari et al [30]. As such, the lower bound of the yield and gain of single photon state denoted respectively as ($Y_1$) and ($Q_1$) in [9], are given as :

$$Y_1 \geq Y_1^L = \frac{\mu}{\mu v - v^2}\left(Q_v e^v - Q_\mu e^\mu \frac{v^2}{\mu^2} - \frac{\mu^2 - v^2}{\mu^2} Y_0\right) \quad (6)$$

$$Q_1 \geq Q_1^L = \frac{\mu^2 e^{-\mu}}{\mu v - v^2}\left(Q_v e^v - Q_\mu e^\mu \frac{v^2}{\mu^2} - \frac{\mu^2 - v^2}{\mu^2} Y_0\right) \quad (7)$$

Similar to the derivation of Weak+Vacum decoy state in [9], taking $v_2 = 0$ and replacing gain from second decoy state $Q_{v2}$ with $Y_0$ and its corresponding error rate with $e_0$ in Eq 10 and 11 of [30] we obtain the lower bound of double photon yield ($Y_2^L$) and gain ($Q_2^L$) for the case of Weak+Vacuum decoy state as :

$$Y_2 \geq Y_2^L = \frac{2\mu\left(Q_v e^v - \frac{v^3}{\mu^3}Q_\mu e^\mu - \frac{\mu^3 - v^3}{\mu^3} Y_0 - \frac{v\mu^2 - v^3}{\mu^2} Y_1^U\right)}{v^2\mu - v^3} \quad (8)$$

$$Q_2 \geq Q_2^L = \frac{\mu^3 e^{-\mu}\left(Q_v e^v - \frac{v^3}{\mu^3}Q_\mu e^\mu - \frac{\mu^3 - v^3}{\mu^3}Y_0 - \frac{v\mu^2 - v^3}{\mu^2}Y_1^U\right)}{v^2\mu - v^3} \quad (9)$$

where $Y_1^U$ is given by :

$$Y_1^U = \frac{(2Q_v e^v - 2Y_0 - Y_2^\infty v^2)}{2v} \quad (10)$$

Note that the $Y_2^\infty$ is the double photon yield from the case of infinite decoy state.
Doing the same for Eq 15 and 16 of [30], the upper bound of single and double photon error rate denoted as $e_1^U$ and $e_2^U$ respectively is then given as :

$$e_1 \leq e_1^U = \frac{E_v Q_v e^v \mu^2 - E_\mu Q_\mu e^\mu v^2 - e_0 Y_0(\mu^2 - v^2)}{Y_1^L(v\mu^2 - \mu v^2)} \quad (11)$$

$$e_2 \leq e_2^U = \frac{2(E_v Q_v e^v \mu - E_\mu Q_\mu e^\mu v - e_0 Y_0(\mu - v))}{Y_2^L(\mu v^2 - v\mu^2)} \quad (12)$$

The resulted $Y_2$, $Q_2$, $e_1$ and $e_2$ together with $Y_1$ and $Q_1$ can now directly be plugged into the secure key rate formula in Eq 28.

**B. The case for $R_{12}$**

Similar to the case of $R_{1+2}$, taking $v_2 = 0$ and substituting the gain from the second decoy state $Q_{v2}$ with $Y_0$ and its corresponding error rate with $e_0$, we rewrite the lower bound $(Y1 + Y2)^L$ in Eq. 19 of [30] for the case of "weak+vacuum", which is now given as :

$$(Y1 + Y2)^L = \frac{\mu^3 e^v Q_v - (\mu^3 - v^3)Y_0 - v^3 Q_\mu e^\mu + \left(v^3\mu - \frac{1}{2}v^3\mu^2\right)Y_1^L}{\mu^3\left(v - \frac{1}{2}\frac{v^3}{\mu}\right)} \quad (13)$$

where $Y_1^L$ is from Eq. 7.

The lower bound of effective gain $Q_{12}^L(\mu)$ and upper bound of effective error rate $\varepsilon^U$ is given by [30] as :

$$Q_{12}^L(\mu) = \left[\frac{(Y1+Y2)^L}{2}\mu^2 + \left(Y_1^L \mu - \frac{Y_1^L \mu^2}{2}\right)\right]e^{-\mu} \quad (14)$$

$$\varepsilon^U = \frac{E_\mu Q_\mu - e_0 Y_0 e^{-\mu}}{Q_{12}^L} \tag{15}$$

The effective gain ($Q_{12}^L(\mu)$) and error rate ($\varepsilon^U$) can be plugged into the following Eq 29 for the lower bound of key generation rate

## 2.3 One Decoy State

The case of one decoy state is similar to the weak+vacuum decoy state except that Bob and Alice do not know their $Y_0$ precisely [9]. Hence they have to estimate the upper bound ($Y_0^U$) which can be imported directly from [9] and is given as :

$$Y_0 \leq Y_0^U = \frac{E_\mu Q_\mu e^\mu}{e_0} \tag{16}$$

From here, we describe the case of $R_{1+2}$ in subsection A and $R_{12}$ in subsection B.

### A. One Decoy State using $R_{1+2}$

Substituting Eq. 16 into Eq. 6 and Eq. 7, we obtain the yield and gain of single photon state denoted as ($Y_1$) and ($Q_1$) as :

$$Y_1 \geq Y_1^L = \frac{\mu}{\mu\nu - \nu^2} \left( Q_\nu e^\nu - Q_\mu e^\mu \frac{\nu^2}{\mu^2} - E_\mu Q_\mu e^\mu \frac{\mu^2 - \nu^2}{e_0 \mu^2} \right) \tag{17}$$

$$Q_1 \geq Q_1^L = \frac{\mu^2 e^{-\mu}}{\mu\nu - \nu^2} \left( Q_\nu e^\nu - Q_\mu e^\mu \frac{\nu^2}{\mu^2} - E_\mu Q_\mu e^\mu \frac{\mu^2 - \nu^2}{e_0 \mu^2} \right) \tag{18}$$

Similarly, replacing Eq. 16 into Eq.8 and Eq.9, we obtain the yield and gain of the double photon state denoted respectively as ($Y_2$) and ($Q_2$), as :

$$Y_2 \geq Y_2^L = \frac{2\mu \left( Q_\nu e^\nu - Q_\mu e^\mu \frac{\nu^3}{\mu^3} - E_\mu Q_\mu e^\mu \frac{\mu^3 - \nu^3}{e_0 \mu^3} - \frac{\nu\mu^2 - \nu^3}{\mu^2} Y_1^U \right)}{\nu^2 \mu - \nu^3} \tag{19}$$

$$Q_2 \geq Q_2^L = \frac{\mu^3 e^{-\mu} \left( Q_\nu e^\nu - Q_\mu e^\mu \frac{\nu^3}{\mu^3} - E_\mu Q_\mu e^\mu \frac{\mu^3 - \nu^3}{e_0 \mu^3} - \frac{\nu\mu^2 - \nu^3}{\mu^2} Y_1^U \right)}{\nu^2 \mu - \nu^3} \tag{20}$$

where $Y_1^U$ is given by :

$$Y_1^U = \frac{(2Q_\nu e^\nu - 2Y_0^L - Y_2^\infty \nu^2)}{2\nu} \tag{21}$$

The $Y_0^L$ in Eq.20 can be obtained following Eq 18 from [9], given as :

$$Y_0^L = max\left\{\frac{(v_1 Q_{v2} e^{v2} - v_2 Q_{v1} e^{v1})}{v_1 - v_2}, 0\right\} \quad (22)$$

where if we take $v2 = 0$ and $Q_{v2} = 0$, we obtain $Y_0^L = 0$.
The upper bound of single and double photon error rate $e_1^U$ and $e_2^U$ is given by :

$$e_1 \le e_1^U = \frac{E_v Q_v e^v \mu^2 - E_\mu Q_\mu e^\mu v^2 - e_0 Y_0^L (\mu^2 - v^2)}{Y_1^L (v\mu^2 - \mu v^2)} \quad (23)$$

$$e_2 \le e_2^U = \frac{2(E_v Q_v e^v \mu - E_\mu Q_\mu e^\mu v - e_0 Y_0^L (\mu - v))}{Y_2^L (\mu v^2 - v\mu^2)} \quad (24)$$

The resulted $Y_2$, $Q_2$, $e_1$ and $e_2$ together with $Y_1$ and $Q_1$ can now directly be plugged into the key rate formula in Eq 28.

## B. One decoy state using $R_{12}$

Similar to the case of $R_{1+2}$, replacing the upper bound of $Y_0^U$ in Eq.16 into Eq.13, we obtain the lower bound $(Y1 + Y2)^L$ as :

$$(Y1 + Y2)^L = \frac{\mu^3 Q_v e^v - (\mu^3 - v^3)\frac{E_\mu Q_\mu e^\mu}{e_0} - v^3 Q_\mu e^\mu + \left(v^3 \mu - \frac{1}{2}v^3 \mu^2\right) Y_1^L}{\mu^3 (v - \frac{1}{2}\frac{v^3}{\mu})} \quad (25)$$

`
where $Y_1^L$ is from Eq. 6.

The lower bound of effective gain $Q_{12}^L(\mu)$ is now given as :

$$Q_{12}^L(\mu) = \left[\frac{(Y1 + Y2)^L}{2}\mu^2 + (Y_1^L \mu - \frac{Y_1^L \mu^2}{2})\right] e^{-\mu} \quad (26)$$

where the $(Y1 + Y2)^L$ and $Y_1^L$ is from Eq.25 and Eq.6 respectively.
Taking $Y_0 = 0$ in Eq.15, then the upper bound of effective error rate $\varepsilon^U$ is now given as :

$$\varepsilon^U = \frac{E_\mu Q_\mu}{Q_{12}^L} \quad (27)$$

The effective gain $(Q_{12}^L(\mu))$ and error rate $(\varepsilon^U)$ can be plugged into the following Eq 29 for the lower bound of key generation rate

## 2.4 The Secure Key Rate

The lower bound of the secure key rate for the case of $R_{1+2}$ and $R_{12}$ is respectively given by [30] as :

$$R_{1+2} \geq R_{1+2}^L = -Q_\mu f(E_\mu)H(E_\mu) + \sum_{i=1}^{2} Q_i[1-\tau(e_i)] \qquad (28)$$

$$R_{12} \geq R_{12}^L = -Q_\mu f(E_\mu)H(E_\mu) + Q_{12}^L[1-\tau(\varepsilon^U)] \qquad (29)$$

where
$H(E_\mu)$ is the binary Shannon Entrophy and is given by

$$H(E_\mu) = -E_\mu \log_2(E_\mu) - (1-E_\mu)\log_2(1-E_\mu) \qquad (30)$$

and $\tau(e)$ as

$$\tau(e) = \log_2(1+4e-4e^2) \text{ for } e < \tfrac{1}{2} \text{ and } \tau(e) = 1 \text{ if } e \geq \tfrac{1}{2} \qquad (31)$$

## 3. RESULTS AND DISCUSSION

In our numerical analysis, we have made use of reliable data obtained from GYS experiment [31]. Using $f(E_\mu)$ so as to match with [30] and the value of $Y_0, \eta_{Bob}$ and $e_{detector}$ from [31], we solved optimal μ and ν numerically and obtain maximum secure key rate for every distance until it hits zero. The last distance before secure key rate hits zero is treated as the maximum secure distance. We have also conducted numerical simulation against the case of without decoy state as well as the case of theoretical infinite decoy state as a base comparison to estimate how well the proposed scheme would performs. For the case of without decoy state, we based on the one in [26]. The result from numerical simulation is depicted in FIG 1.

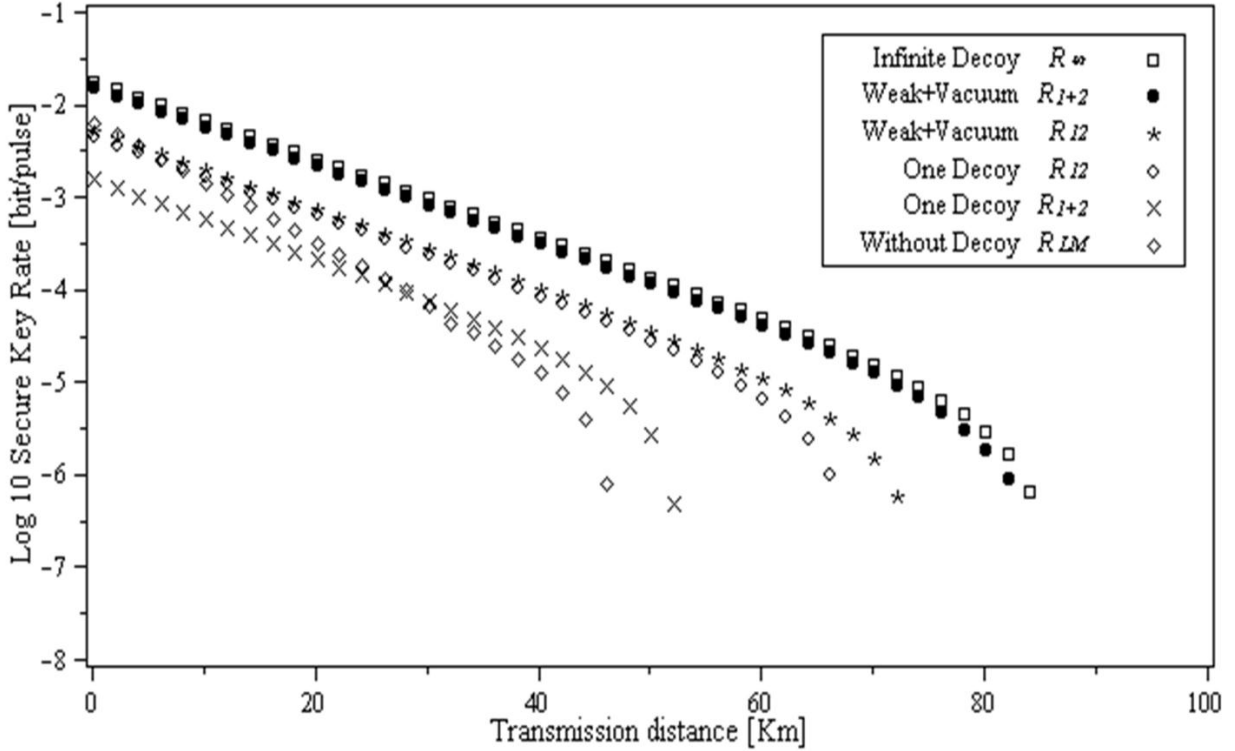

FIG. 1 A plot of key generation rate against transmission distance from the result of numerical simulation for all the six cases mentioned in the text (Infinite decoy state ($R_\infty$), weak+vacuum decoy state ($R_{1+2}$ and $R_{12}$), one decoy state ( derived from $R_{1+2}$ and $R_{12}$) and without decoy state ($R_{LM}$) . The system parameters are from GYS [31].

It can easily be seen that for the case of $R_{1+2}$, the maximum secure distance is extended by almost double while for the case of $R_{12}$ although not as good as $R_{1+2}$, was able to extend by almost two third the maximum distance of the case of without decoy state. Comparing against the theoretical infinite case, $R_{1+2}$ performs very well close to the limit while $R_{12}$ was a bit shorter by around 10 km. Another aspect of the $R_{12}$ result is that the secure key rate is a bit low with a consistent gap until near to 60 km. The results match well with one in [30] and indicate the practicality of the weak+vacuum decoy state in extending the maximum secure distance of a two way protocol specifically the LM05.

For the case of one decoy state, the one derived for the $R_{12}$ formula had surprisingly performed better than the one derived from the $R_{1+2}$ and was quite close to the case of $R_{12}$ (weak+vacuum). Although the maximum secure distance of one decoy state using $R_{1+2}$ formula was slightly higher than the case of without decoy state, the secure key rate prior to 30 km was unexpectedly worse. This indicates the impracticality of the proposed bound.

## 4. CONCLUSION AND FUTURE WORKS

We have derived the bounds for specific case of weak+vacum decoy state and one decoy state with a two way Quantum Key Distribution protocol namely the LM05. The numerical simulation results for the weak+vacuum decoy state was quite encouraging given that the maximum secure distance was extended by almost double and was very well close to the case of maximum theoretical limit. The result also indicates the practicality of one of the proposed bound for the case of one decoy state that is when the yield calculation of single and double photon contribution was lumped. With all the practical bounds ready, it is very much interesting to see the proposed decoy state extension in action. A simple extension to the source as well as modification of the programming on our previously developed free space based LM05 QKD system in [33] is sufficient enough to accommodate the proposed decoy state extension.